\documentclass[12pt]{article}
 
\usepackage{graphicx}
\usepackage{amsfonts}
\usepackage{amsthm}
\usepackage[top=2.75cm, bottom=3cm, left=2.75cm, right=2.75cm]{geometry}

\newcommand{\beq}{\begin{equation}} 
\newcommand{\eeq}[1]{\label{#1}\end{equation}}
\newcommand{\ber}{\begin{eqnarray}} 
\newcommand{\eer}[1]{\label{#1}\end{eqnarray}} 
\newcommand{\re}[1]{(\ref{#1})} 

\newcommand{\ft}[2]{{\textstyle\frac{#1}{#2}}}
\newcommand{\nll}{N\!=\!(1,1)}
\newcommand{\nZZ}{N\!=\!(2,2)}
\newcommand{\ff}{\mathbb{F}}
\newcommand{\ffb}{{\bar\mathbb{F}}}
\newcommand{\fft}{{\tilde\mathbb{F}}}
\newcommand{\fftb}{{\bar{\tilde\mathbb{F}}}}

\newcommand{\vv}[1]{\mathbb{V}^{#1}}
\newcommand{\vvb}[1]{\bar{\mathbb{V}}^{#1}}
\newcommand{\vvt}[1]{\tilde{\mathbb{V}}^{#1}}
\newcommand{\vvtb}[1]{\bar{\tilde{\mathbb{V}}}^{#1}}

\newcommand{\dt}[1]{\hat{d}^{#1}}
\newcommand{\dq}[1]{\hat{q}^{#1}}
\newcommand{\jj}[3]{{J}_{#1}{}^{#2}{}_{#3}}
\newcommand{\bbD}[1]{\mathbb{D}_{#1}}
\newcommand{\bbDB}[1]{\bar{\mathbb{D}}_{#1}}
\newcommand{\bbG}[1]{\mathbb{G}_{#1}}
\newcommand{\bbGB}[1]{\bar{\mathbb{G}}_{#1}}
\newcommand{\bbX}[1]{\mathbb{X}_{#1}}
\newcommand{\bbXB}[1]{\bar{\mathbb{X}}_{#1}}

\newcommand{\hee}[1]{\Xi^1_{#1}}
\newcommand{\hcc}[1]{\Xi^2_{#1}}

\newcommand{\hcd}[1]{\nabla_{#1}}
\newcommand{\hPh}[1]{\varphi^{#1}}
\newcommand{\hph}{\phi}
\newcommand{\hch}{\chi}
\newcommand{\hphb}{\bar{\phi}}
\newcommand{\hchb}{\bar{\chi}}
\newcommand{\redQp}[2]{ \jj{+}{#1}{#2} \hcd{+} \hPh{#2} + \hee{+} \jj{-}{#1}{#2} k^{#2} + \hcc{+} \Pi^{#1}{}_{#2}k^{#2}}
\newcommand{\redQm}[2]{ \jj{-}{#1}{#2} \hcd{-}\hPh{#2} + \hee{-} \jj{+}{#1}{#2} k^{#2} + \hcc{-} \Pi^{#1}{}_{#2}k^{#2}}

\newcommand{\aleq}{&\!\!=\!\!&}
\newcommand{\nn}{\nonumber}
\newcommand{\kah}{K\"ahler~}
\newcommand{\pa}[1]{\partial_{#1}}
\newcommand{\lam}{\Lambda}
\newcommand{\G}{\Gamma}
\newcommand{\lamb}{\bar\Lambda}
\newcommand{\lamt}{\tilde\Lambda}
\newcommand{\lamtb}{\bar{\tilde{\Lambda}}}
\newcommand{\etc}{\textit{etc}}
\newcommand{\eg}{\textit{e.g.},}
\begin{document}
\renewcommand{\theequation}{\thesection.\arabic{equation}}  
\setcounter{page}{0}
\thispagestyle{empty}

\begin{flushright} \small
UUITP-08/07 \\ HIP-2007-28/TH \\NORDITA-2007-15\\  YITP-SB-07-18 \\ 
\end{flushright}
\smallskip
\begin{center} \LARGE
{\bf New $\nZZ$ vector multiplets}
 \\[12mm] \normalsize
{\bf Ulf~Lindstr\"om$^{a,b,c}$, Martin Ro\v cek$^{d}$, Itai Ryb$^{d}$,\\
Rikard von Unge$^{e}$, and Maxim Zabzine$^{a}$} \\[8mm]
{\small\it
$^a$Department of Theoretical Physics 
Uppsala University, \\ Box 803, SE-751 08 Uppsala, Sweden \\
~\\
$^b$HIP-Helsinki Institute of Physics, University of Helsinki,\\
P.O. Box 64 FIN-00014  Suomi-Finland\\
~\\
$^c$NORDITA, Roslagstullsbacken 23,\\
SE-10691 Stockholm, Sweden\\
~\\
$^d$C.N.Yang Institute for Theoretical Physics, Stony Brook University, \\
Stony Brook, NY 11794-3840,USA\\
~\\
$^{e}$Institute for Theoretical Physics, Masaryk University, \\ 
61137 Brno, Czech Republic \\~\\}
\end{center}
\vspace{10mm}
\centerline{\bf\large Abstract} 
\bigskip
\noindent  
We introduce two new $\nZZ$ vector multiplets that couple naturally 
to generalized K\"ahler geometries. We describe their kinetic actions as well as
their matter couplings both in $\nZZ$ and $\nll$ superspace.
\eject
\normalsize

%\addtocontents{toc}
%\ofcontents

%\end{titlepage}

\eject

\section{Introduction}
Generalized \kah geometry has aroused considerable interest both among string theorists 
and mathematicians, \eg\ \cite{Lindstrom:2004iw,gualtieri,many}. 
Recently, several groups have tried to construct quotients
\cite{lin_tolman, hu, bcg, Merrell:2006py}; 
however, it is unclear how general or useful the various proposals are.
Experience has shown that supersymmetric $\sigma$-models are often
a helpful guide to finding the correct geometric concepts and framework for
quotient constructions \cite{lr,hklr}. In this paper, we take the first step in this direction;
further results will be presented in \cite{LRRvUZ}.

The basic inspiration for our work is the interesting duality found in 
\cite{Grisaru:1997ep,Bogaerts:1999jc}. As was shown in \cite{hklr,Rocek:1991ps},
T-dualities arise when one gauges an isometry, and then constrains
the field-strength of the corresponding gauge multiplet to vanish. Here
we address the question: what are the gauge multiplets corresponding to the
duality introduced in \cite{Grisaru:1997ep,Bogaerts:1999jc}?

In section 2, we analyze the types of isometries that arise on generalized \kah
geometries which are suitable for gauging, and describe the corresponding multiplets
in $\nZZ$ superspace. In addition to the usual multiplets with chiral or twisted chiral
gauge parameters, we find two new multiplets: one with semichiral gauge parameters,
which we call the semichiral gauge multiplet, and one with a pair of gauge parameters,
one chiral and one twisted chiral; the last has more gauge-invariant
components than other multiplets, and hence we call it the large vector multiplet. 

In section 3, we describe the $\nll$ superspace content of these mulitplets; this
exposes their physical content.  We describe both multiplets
and their couplings to matter, and discuss possible gauge actions for them.
The component content of the various $\nll$ multiplets that arise is well known and can be found in \cite{thebook}.

Throughout this paper we follow the conventions of \cite{Lindstrom:2005zr}.
\section{Generalized K\"ahler geometry: $\nZZ$ superspace}
\setcounter{equation}{0}
Generalized K\"ahler geometry (GKG) arises naturally as the target space of $\nZZ$ 
supersymmetric $\sigma$-models. As shown in \cite{Lindstrom:2005zr}, such 
$\sigma$-models always admit a local description in $\nZZ$ superspace in terms of
complex chiral superfields $\phi$, twisted chiral superfields $\chi$ and semichiral 
superfields $\bbX{L},\bbX{R}$ \cite{Buscher:1987uw}. These models have also
been considered in $\nll$ superspace \cite{Gates:1984nk,Kapustin:2006ic}.

These geometries may admit a variety of holomorphic isometries that can be gauged 
by different kinds of vector multiplets. We now itemize the basic types of isometries.

\subsection{Isometries}

The simplest isometries act on purely \kah submanifolds of the generalized \kah geometry, 
that is only on the chiral superfields $\phi$ or the twisted chiral superfields $\chi$; 
for a single $U(1)$ isometry away from a fixed point, we may choose coordinates so that
 the Killing vectors take the form:
\begin{equation}
\label{eq::kaki}
k_\phi = i (\pa{\phi}-\pa{\bar\phi}) ~~,~~~~ k_\chi = i (\pa{\chi}-\pa{\bar\chi}) ~~.
\end{equation}

In \cite{Grisaru:1997ep,Bogaerts:1999jc}, new isometries that mix chiral and 
twisted chiral superfields or act on semichiral superfields were discovered; 
we may take them to act as 
\begin{eqnarray}
k_{\phi \chi}\aleq i(\pa{\phi}-\pa{\bar\phi}-\pa{\chi}+\pa{\bar\chi}) ~,\\
k_{LR}\aleq i(\pa{L}-\pa{\bar{L}}-\pa{R}+\pa{\bar{R}}) ~,
\end{eqnarray}
where $\pa{L} = \frac{\pa{}}{\pa{}\bbX{L}}$, \etc. 
One might imagine more general isometries that act along an arbitrary vector
field; however, compatibility with the constraints on the superfields 
(chiral and twisted chiral superfields are automatically semichiral but not
vice-versa) allows us to restrict to the cases above; in particular, 
if the vector field has a component along $k_\phi,k_\chi$ or $k_{\phi\chi}$, 
we can (locally) redefine $\bbX{}$ to
eliminate any component along $k_{LR}$.

A general Lagrange density in $\nZZ$ superspace has the form:
\begin{equation}
\label{eq::KAHLER_K}
K=K(\phi,\bar{\phi},\chi,\bar{\chi}, \bbX{L},\bbXB{L},\bbX{R},\bbXB{R} )
\end{equation}
For the four isometries listed above the corresponding invariant Lagrange densities 
are\footnote{Generally, isometries may leave the Lagrange density invariant only up 
to a (generalized) \kah transformation \cite{Hull:1985pq,Lindstrom:2005zr}, 
but as our interest here is the structure of the vector multiplet, 
we are free to choose the simplest situation.}:
\begin{eqnarray}
\label{eq::invlag1}&&k_\phi \, K(\phi+\bar{\phi},\chi,
\bar{\chi}, \bbX{L},\bbXB{L},\bbX{R},\bbXB{R} )=0\\[2mm]
\label{eq::invlag2}&&k_\chi \, K(\phi,\bar{\phi},\chi
+\bar{\chi}, \bbX{L},\bbXB{L},\bbX{R},\bbXB{R} )=0\\[2mm]
\label{eq::invlag3}&&k_{\phi\chi} \, K(\phi+
\bar{\phi},\chi+\bar{\chi} ,i(\phi-\bar\phi+\chi-\bar\chi), 
\bbX{L},\bbXB{L}, \bbX{R},\bbXB{R} ) = 0\\[2mm]
\label{eq::invlag4}&&k_{LR} \, K(\phi,\bar{\phi},\chi,
\bar{\chi}, \bbX{L}+\bbXB{L},\bbX{R}+\bbXB{R}, i(\bbX{L}
-\bbXB{L} +\bbX{R}-\bbXB{R}) )= 0
\end{eqnarray}

In general, the isometries act on the coordinates with some constant parameter $\lambda$:
\begin{equation}
\label{eq::glokz}
\delta z = [\lambda k,z]~~,
\end{equation}
where $z$ is any of the coordinates $\phi,\chi,\bbX{L},\bbX{R}$, etc.

\subsection{Gauging and Vector Multiplets}
We now promote the isometries to local gauge symmetries:
the constant transformation parameter $\lambda$ of (\ref{eq::glokz})
becomes a local parameter $\lam$ that obeys the appropriate constraints.
\begin{eqnarray}
\label{eq::loctrans} \delta_g \phi = i\lam &\Rightarrow&
\bbDB{\pm} \lam = 0 \nn\\
 \delta_g \bar\phi = -i\lamb
&\Rightarrow& \bbD{\pm} \lamb = 0 \nn\\
 \delta_g \chi = i\lamt
&\Rightarrow& \bbDB{+} \lamt = \bbD{-} \lamt = 0 \nn \\
 \delta_g \bar\chi = -i\lamtb &\Rightarrow&
\bbD{+} \lamtb =\bbDB{-} \lamtb = 0 \nn \\
 \delta_g \bbX{L} = i \lam_L &\Rightarrow& \bbDB{+}
\lam_L = 0 \nn \\
\delta_g \bbX{R} = i \lam_R &\Rightarrow& \bbDB{-}
\lam_R = 0 \nn \\
 \delta_g \bbXB{L} = -i \lamb_L &\Rightarrow&
\bbD{+}\lamb_L = 0 \nn \\
 \delta_g \bbXB{R} = -i \lam_R &\Rightarrow&
\bbD{-}\lamb_R = 0~.
\end{eqnarray}
To ensure the invariance of the Lagrange densities (\ref{eq::invlag1}-\ref{eq::invlag4})
under the local transformations (\ref{eq::loctrans}), we introduce 
the appropriate vector multiplets. For the isometries 
(\ref{eq::invlag1},\ref{eq::invlag2}) these give the well known transformation 
properties for the usual (un)twisted vector multiplets:
\begin{eqnarray}
\delta_g V^\phi = i(\lamb-\lam) &\Rightarrow& \delta_g
(\phi+\bar\phi+V^\phi) = 0\nn\\
\delta_g V^\chi = i(\lamtb-\lamt) &\Rightarrow&
\delta_g (\chi+\bar\chi+V^\chi) = 0~,
\end{eqnarray}
whereas for generalized \kah transformations we need to add \textit{triplets} 
of vector multiplets. 

For the the semichiral isometry $k_{LR}$, we introduce the vector multiplets:
\begin{eqnarray}\label{dsemi}
\delta_g \vv{L}= i(\lamb_L-\lam_L) &\Rightarrow&
\delta_g (\bbX{L}+\bbXB{L}+\vv{L}) = 0\nn\\
\delta_g \vv{R}= i(\lamb_R-\lam_R) &\Rightarrow&
\delta_g (\bbX{R}+\bbXB{R}+\vv{R}) = 0\nn\\
\delta_g \vv{\prime} = \lam_L+\lamb_L+\lam_R+\lamb_R
&\Rightarrow& \delta_g (i(\bbX{L}-\bbXB{L}+\bbX{R}-\bbXB{R})+\vv{\prime}) = 0~.
\end{eqnarray}
We refer to this multiplet as the semichiral vector multiplet.

For the $k_{\phi\chi}$ isometry we introduce the 
vector multiplets
\begin{eqnarray}
\delta_g V^\phi = i(\lamb-\lam) &\Rightarrow& \delta_g (\phi+\bar\phi+V^\phi) = 0\nn\\
\delta_g V^\chi = i(\lamtb-\lamt) &\Rightarrow& \delta_g (\chi+\bar\chi+V^\chi) = 0\nn\\
\delta_g V' = \lam+\lamb+\lamt+\lamtb &\Rightarrow& 
\delta_g (i(\phi-\bar\phi+\chi-\bar\chi)+V') = 0~,
\end{eqnarray}
and refer to this multiplet as the large vector multiplet due to the large number of
gauge-invariant components that comprise it.
\subsection{$\nZZ$ field-strengths}
We now construct the $\nZZ$ gauge invariant field-strengths for the 
various multiplets introduced above.
\subsubsection{The known field-strengths}
The field-strengths for the usual vector multiplets are well known:
\begin{eqnarray}
\nonumber &&\tilde{W} =i\, \bbD{-}\bbDB{+} V^\phi~~,~~~~
\bar{\tilde{W}} =i\, \bbDB{-}\bbD{+} V^\phi~, \\
&&W =i\, \bbDB{-}\bbDB{+} V^\chi ~~,~~~~
\bar{W} =i\, \bbD{-}\bbD{+} V^\chi ~. 
\end{eqnarray}
Note that $\tilde{W}$, the field-strength for the chiral isometry is twisted chiral whereas $W$, the field-strength for the twisted chiral isometry, is chiral.

\subsubsection{Semichiral field-strengths}
To find the gauge-invariant field-strengths for the vector multiplet 
that gauges the semichiral isometry it is useful to introduce the 
complex combinations:
\begin{eqnarray}
\vv{} = \frac{1}{2}(\vv{\prime}+i(\vv{L}+\vv{R})) &\Rightarrow& 
\delta_g \vv{} = \lam_L + \lam_R~, \nn\\
\vvt{} = \frac{1}{2}(\vv{\prime}+i(\vv{L}-\vv{R})) &\Rightarrow& 
\delta_g \vvt{} = \lam_L + \lamb_R ~.
\end{eqnarray}
Then the following complex field-strengths are gauge invariant:
\begin{eqnarray}\label{semiF}
\ff \aleq \bbDB{+}\bbDB{-}\vv{} ~~,~~~~ 
\ffb = - \bbD{+}\bbD{-}\vvb{}~,\nn \\ 
\fft \aleq \bbDB{+}\bbD{-}\vvt{} ~~,~~~~ 
\fftb = -\bbD{+}\bbDB{-}\vvtb{} ~,
\end{eqnarray}
where $\ff$ is chiral and $\fft$ is twisted chiral. 

\subsubsection{Large Vector Multiplet field-strengths}
As above it is useful to introduce the complex potentials:
\begin{eqnarray}
V = \frac{1}{2} [V^\prime +i(V^\phi+V^\chi)] 
&\Rightarrow& \delta_g V = \lam+\lamt~, \nn\\ 
\tilde{V} = \frac{1}{2} [V^\prime +i(V^\phi-V^\chi)] &\Rightarrow& \delta_g 
\tilde{V} = \lam+\lamtb ~.
\end{eqnarray}
Because $(\lamt)\lam $ are (twisted)chiral respectively, the following 
complex spinor field-strengths are gauge invariant:
\begin{eqnarray}
\bbG{+} \aleq \bbDB{+} V ~~,~~~~ \bbGB{+} = \bbD{+} \bar{V}~,\nn\\
\bbG{-} \aleq \bbDB{-} \tilde{V} ~~,~~~~ \bbGB{-} = \bbD{-} \bar{\tilde{V}}~.
\end{eqnarray}

The higher dimension field-strengths can all be constructed from these spinor field-strengths:
\begin{eqnarray}\label{wb}
\nn W\aleq -i \bbDB{+}\bbDB{-} V^\chi = 
\bbDB{+} \bbG{-} +\bbDB{-} \bbG{+} \\ 
\nn \bar{W}	\aleq -i \bbD{+}\bbD{-} V^\chi = 
-(\bbD{+} \bbGB{-} +\bbD{-} \bbGB{+}) \\ 
\nn \tilde{W} \aleq -i \bbD{+}\bbDB{-} V^\phi =
 \bbDB{+} \bbGB{-} +\bbD{-} \bbG{+} \\
\nn\bar{\tilde{W}} \aleq -i \bbDB{+}\bbD{-} V^\phi =
-(\bbD{+} \bbG{-} +\bbDB{-} \bbGB{+}) \\
\nn B \aleq - \bbDB{+}\bbDB{-} (V^\prime + i V^\phi) =
\bbDB{-} \bbG{+} - \bbDB{+} \bbG{-} \\
\nn\bar{B} \aleq \bbD{+}\bbD{-} (V^\prime - i V^\phi) = 
-(\bbD{-} \bbGB{+} - \bbD{+} \bbGB{-} )\\
\nn \tilde{B} \aleq - \bbD{+}\bbDB{-} (V^\prime - i V^\chi) = 
\bbD{-} \bbG{+} - \bbDB{+} \bbGB{-} \\
\bar{\tilde{B}} \aleq \bbDB{+}\bbD{-} (V^\prime + i V^\chi) =
-(\bbDB{-} \bbGB{+} - \bbD{+} \bbG{-} ) ~;
\end{eqnarray}
the chirality properties of these field-strengths are summarized below:
\begin{equation}
\begin{array}{|c||c|}
\hline
\hbox{Field-strength} & \hbox{Property} \\
\hline \hline
W ,B & \hbox{chiral} \\
\bar{W},\bar{B} & \hbox{anti-chiral}\\
\tilde{W},\tilde{B} & \hbox{twisted chiral}\\
\bar{\tilde{W}},\bar{\tilde{B}} & \hbox{anti-twisted chiral}\\
\hline
\end{array}
\end{equation}
\section{Gauge multiplets in $\nll$ superspace}
\setcounter{equation}{0}
To reveal the physical content of the gauge multiplets, we could go to components, but it
is simpler and more informative to go to $\nll$ superspace. We expect to find
spinor gauge connections and unconstrained superfields. As mentioned in the introduction, the component content of various $\nll$ multiplets can be found in \cite{thebook}.

The procedure for going to $\nll$ components is well-known; for a convenient
review, see \cite{Lindstrom:2005zr}. We write the $\nZZ$ derivatives $\bbD{\pm}$ 
and their complex conjugates $\bbDB{\pm}$ in terms of real $\nll$ derivatives 
$D_\pm$ and the generators $Q_\pm$ of the nonmanifest supersymmetries,
\begin{equation}
\bbD{\pm} = \frac{1}{2}(D_\pm - iQ_\pm) ~~,~~~~ \bbDB{\pm} 
= \frac{1}{2} (D_\pm + iQ_\pm)~,
\end{equation}
and $\nll$ components of an unconstrained superfield $\Psi$ as $\Psi|=\phi$,
$Q_\pm\Psi|=\psi_\pm$, and $Q_+Q_-\Psi|=F$. 
\subsection{The semichiral vector multiplet}
We first identify the $\nll$ components of the semichiral vector multiplet,
and then describe various couplings to matter.
\subsubsection{$\nll$ components of the gauge multiplet}
We can find all the $\nll$ components of the semichiral gauge multiplet from the 
field strengths (\ref{semiF}) except for the spinor connections $\G_\pm$. The only 
linear combination of the gauge parameters $\lam_R,\lam_L$ that does not enter algebraically in (\ref{dsemi}) is $(\lam_L+\lamb_L-\lam_R-\lamb_R)$, and hence the connections must transform as:
\begin{equation}
\delta_g \G_\pm = \left.\frac{1}{4} D_\pm (\lam_L+\lamb_L-\lam_R-\lamb_R)\right|~.
\end{equation}
This allows us to determine the connections as:
\beq
\G_+=\left.\left(\frac12Q_+\vv{L}-\frac14D_+\vv{\prime}\right)\right|~~,~~~
\G_-=-\left.\left(\frac12Q_-\vv{R}-\frac14D_-\vv{\prime}\right)\right|~,
\eeq{semicon}
where the $D_\pm$ terms vanish in Wess-Zumino gauge. 
The gauge-invariant component fields are just
the projections of the $\nZZ$ field-strengths (\ref{semiF}) and the field-strength of the
connection $\G_\pm$:
\beq
f=i(D_+\G_-+D_-\G_+)~.
\eeq{fsemi}
These are not all independent--they obey the Bianchi identity:
\beq
f= i \left.\left(\ff-\ffb+\fft-\fftb\right)\right|~.
\eeq{bianch1}
Thus this gauge multiplet is described by an $\nll$ gauge multiplet 
and three real unconstrained
$\nll$ scalar superfields:
\beq
\dt1=\!\left.\left(\ff+\ffb\right)\right|~~,~~~\dt2=\!\left.\left(\fft+\fftb\right)\right|
~~,~~~\dt3=\!\left.i\!\left(\ff-\ffb-\fft+\fftb\right)\right|~.
\eeq{ffN11}
Though not essential, the simplest way to find the $\nll$ reduction of
various $\nZZ$ quantities is to go to a Wess-Zumino gauge, that is reducing
the $\nZZ$ gauge parameters to a {\em single} $\nll$ gauge parameter
by gauging away all $\nll$ components with algebraic gauge transformations.
Here this means imposing
\beq
\begin{tabular}{lll}
$\vv{L}|=0~~,~~~$&$ (Q_+\vv{L})|=2\G_+ ~~,~~~$& $(Q_-\vv{L})|= 0~~,$\\
$\vv{R}|=0~~,~~~$&$ (Q_+\vv{R})|=0 ~~,~~~$& $(Q_-\vv{R})|= -2\G_-~~,$ \\
$\vv{\prime}|=0~~,~~~$&$ (Q_+\vv{\prime})|=0~~,~~~$& $(Q_-\vv{\prime})|= 0~~,$
\end{tabular}
\eeq{WZsemiV}
on the gauge multiplet and
\beq
\lam^L|=\lamb^L|=-\lam^R|=-\lamb^R|~~,~~~
(Q_-\lam^L)|=(Q_-\lamb^L)|=(Q_+\lam^R)|=(Q_+\lamb^R)|=0
\eeq{WZsemilam}
on the gauge parameters.
This leads directly to:
\beq
(Q_+Q_-\vv{L})|=2i(\dt1-\dt2)~~,~~~(Q_+Q_-\vv{R})|=2i(\dt1+\dt2)~~,~~~
(Q_+Q_-\vv{\prime})|=2i\dt3~~.
\eeq{WZsemiQQ}
\subsubsection{Coupling to matter}
We start from the gauged $\nZZ$ Lagrange density:
\begin{equation}
\label{eq::semi_gauged_action}
K_\mathbb{X} = K_\mathbb{X} \left( \bbX{L}+\bbXB{L}+\vv{L} , \bbX{R}+\bbXB{R}+\vv{R} , i(\bbX{L}-\bbXB{L}+\bbX{R}-\bbXB{R})+\vv{\prime}\right)~.
\end{equation}
In the Wess-Zumino gauge defined above, we have
\beq
X_{L(R)}= \bbX{L(R)}|~,
\eeq{semixdef}
and $\nll$ spinor components:
\begin{eqnarray}
(Q_+ \bbX{L})| = i D_+ X_{L} + \G_+ ~&,&~~ (Q_- \bbX{L})| = \psi_-~,\nn \\
(Q_- \bbX{R})| = i D_- X_{R} - \G_- ~&,&~~ (Q_+ \bbX{R})| = \psi_+ ~.
\end{eqnarray}\label{semipsidef}
Then for the tuple $X^i$ and the isometry vector $k^i$ defined as
\begin{eqnarray}
k^i &\!\!\equiv\!\!& k_{\phi \chi} = k_{LR} = (i,-i,-i,i) ~,\nn\\
X^i \aleq (X_{L},\bar X_{L},X_{R},\bar X_{R})~,
\end{eqnarray}
we write the gauge covariant derivative as it appears in \cite{hklr}
\begin{equation}
\hcd{\pm} X^i = D_\pm X^i - \G_\pm k^i.
\end{equation}
We can compute
\begin{eqnarray}
(Q_+ Q_- \bbX{L})| \aleq i D_+ \psi_- + i(\dt{1}-\dt{2})+ \dt{3} \nn\\
(Q_+ Q_- \bbX{R})| \aleq -i D_+ \psi_- + i(\dt{1}+\dt{2})+ \dt{3} ~.
\end{eqnarray}
Using
\begin{eqnarray}
\label{eq::hessann}
\frac{\partial^2 K}{\partial  X^{i} \partial  X^{j}} k^i = 0~~ 
\Rightarrow ~~~\frac{\partial^2 K}{\partial  X^{i}
\partial  X^{j}}D_\pm  X^{i} =\frac{\partial^2 K}{\partial  X^{i}
\partial  X^{j}} \hcd{\pm}  X^{i} ~,\\ \nn
\end{eqnarray}
we obtain the gauged $\nll$ Lagrange density
\beq
E_{ij}\hcd{+}X^i \hcd{-}X^j + K_i L^i{}_\alpha \dt{\alpha}~,
\eeq{semilagn11}
with:
\begin{equation}
L=\left( \begin{array}{rrr}
i&-i&1\\
-i&i&1\\
i&i&1\\
-i&-i&1\\
\end{array}
 \right)~.
\end{equation}
Here $E=\frac{1}{2}(g+B)$ in the reduced Lagrange density is that same as 
for the ungauged $\sigma$-model \cite{Lindstrom:2005zr,Lindstrom:2007qf}.
\subsubsection{The vector multiplet action}
Introducing the notation
\begin{equation}
\mathbb{F}^i \equiv ( \ff,\ffb,\fft,\fftb )~~,~~~~ d^i \equiv ( f,\dt{1},\dt{2},\dt{3})~,
\end{equation}
and using the (twisted)chirality properties
\begin{equation}
\bbDB{\pm} \ff =  \bbD{\pm}\ffb = \bbDB{+} \fft = 
\bbD{-}\fft = \bbD{+}\fftb = \bbDB{-}\fftb = 0 ~,
\end{equation}
we find
\begin{equation}
(Q_\pm \mathbb{F}^i)| = \jj{\pm}{i}{j}\, M^j{}_k (D_\pm\dt{k})~,
\end{equation}
with
\begin{equation}
 M = \frac{1}{4}\left( \begin{array}{rrrr}
-i & 2 & 0 & -i \\
i & 2 & 0 & i \\
-i & 0 & 2 & i \\
i & 0 & 2 & -i \\
\end{array}\right)~~,~~~J_\pm \equiv \hbox{diag}(i,-i,\pm i,\mp i) ~.
\end{equation}
Starting from an $\nZZ$ action:
\begin{equation}
S_\mathbb{X} =  \int d^2 \xi\, D_+ D_- Q_+ Q_- 
\left(a\,\ff\ffb-b \,\fft\fftb \right)
\end{equation}
we write the reduction to $\nll$ in 
terms of the gauge-invariant $\nll$ components $\dt{i}$:
\begin{equation}
S_{\mathbb{X}} = \frac12 \int d^2 \xi\, D_+ D_-\left(D_+ \dt{i}\, D_-\dt{j} \, g_{ij}\right)~,
\eeq{Ssemivec11}
where
\begin{equation}
g = \frac{1}{8} \left( \begin{array}{cccc}
a+b	& 0	& 0 	& a-b \\
0	& 4a	& 0 	& 0 \\
0	& 0	& 4b & 0\\
a-b	& 0	& 0 	& a+b 
\end{array}
\right) ~.
 \end{equation}
To obtain real and positive definite $g$ we require $ab>0$ 
which yields one $\nll$ gauge multiplet and three scalar multiplets. In particular,
when $a=b$, we find the usual diagonal action.

Other gauge-invariant terms are possible; these are general superpotentials
and have the form
\begin{equation}
S_{P} = \int i\bbD{+} \bbD{-} \,P_1 (\ff) + \int i\bbDB{+} \bbDB{-}\,\bar P_1 (\ffb) 
+ \int i\bbD{+} \bbDB{-}\, P_2 (\fft) + \int i\bbDB{+} \bbD{-} \,\bar P_2 (\fftb)   ~,
\end{equation}
where $P$ are holomorphic functions. 
These terms reduce trivially to  give:
\beq
S_{P} = 2\int iD_+ D_- ~\hbox{Re}\!\left(P_1 (\ft12\dt{1}-\ft{i}4(f+\dt{3})) + 
P_2 (\ft12\dt{2}-\ft{i}4(f-\dt{3})) \right).
\eeq{semiP}
Particular examples of such superpotentials include 
mass and Fayet-Iliopoulos terms.

\subsubsection{Linear terms}
To perform T-duality transformations, one gauges an isometry, and then constrains
the field-strength to vanish \cite{hklr,Rocek:1991ps}. We will discuss T-duality for
generalized \kah geometry in detail in \cite{LRRvUZ}; it was introduced 
(without exploring the gauge aspects) in \cite{Grisaru:1997ep,Bogaerts:1999jc}.
Here we describe the $\nZZ$ superspace coupling and its reduction to $\nll$.
We constrain the field-strengths to 
vanish using unconstrained complex Lagrange multiplier superfields $\Psi,\tilde\Psi$
\beq
\mathcal{L}_{linear}= \Psi\ff+\bar\Psi\ffb+\tilde\Psi\fft
+\bar{\tilde\Psi}\fftb~;
\eeq{linpsi}
integrating by parts, we can re-express this in terms of
chiral and twisted chiral Lagrange multipliers 
$\phi=\bbDB+\bbDB-\Psi$, $\chi=\bbDB+\bbD-\tilde\Psi$ to obtain
\begin{equation}\label{semillin}
\mathcal{L}_{linear}= \phi \vv{} + \bar\phi \vvb{} 
+ \chi \vvt{} + \bar\chi \vvtb{}~.
\end{equation}
This reduces to an $\nll$ superspace Lagrange density (up to total derivative terms)
\begin{eqnarray}
\mathcal{L}_{linear}\aleq \phi (i\dt{3}-2\dt{1}+if) 
+ \bar\phi (i\dt{3}+2\dt{1}+if)\nn\\ 
&&+\,\,\chi(i\dt{3}+2\dt{2}-if)+\bar\chi (i\dt{3}-2\dt{2}-if)~,
\end{eqnarray}
where $\phi,\bar\phi,\chi,\bar\chi$ are the obvious $\nll$ projections of the
corresponding $\nZZ$ Lagrange multipliers. When we perform a T-duality
transformation, we add this to the Lagrange density \re{semilagn11}.
\subsection{The Large Vector Multiplet}
We now study the $\nll$ components of the large vector multiplet.
\subsubsection{$\nll$ gauge invariants}
Starting with the eight $\nZZ$ second-order gauge invariants (\ref{wb}), we
descend to $\nll$ superspace and identify the $\nll$
gauge field-strength. 

Imposing the condition that the $\nll$ gauge connection transforms as 
\begin{equation}
\delta_g A_\pm = \frac{1}{4} D_\pm (\lamtb+\lamt-\lamb-\lam)~,
\end{equation}
we find the quantities
\begin{eqnarray}
A_+ \aleq -\left.\left(\frac{1}{4} Q_+(V^\phi - V^\chi )\right)\right| = \left.\left(
\frac{i}{4}Q_+ (\tilde{V}-\bar{\tilde{V}})\right)\right|,\nn \\
A_- \aleq -\left.\left(\frac{1}{4} Q_-(V^\phi + V^\chi )\right)\right| =\left.\left(
\frac{i}{4}Q_- (V-\bar{V})\right)\right|;
\end{eqnarray}
of course, any gauge-invariant spinor may be added to $A_\pm$. 
It is useful to introduce the real and imaginary parts of $\bbG{\pm}$:
\begin{equation}
\Xi^A_{\pm} = \left(\,\left. \hbox{Re} (\bbG{\pm})\right| ,
\left. \hbox{Im} (\bbG{\pm}) \right|\,\right)~.
\end{equation} 
These form a basis for the $\nll$ gauge-invariant spinors. 
The field-strength of the connection $A_\pm$
\begin{equation}
f=i(D_+ A_- + D_- A_+) = i(Q_+ \hcc{-}+Q_- \hcc{+}) 
\end{equation}
is manifestly gauge invariant. The remaining $\nll$ gauge-invariant scalars are:
\begin{eqnarray}
\dq{1}=i( Q_- \hee{+} - Q_+ \hee{-})~,\nn \\
\dq{2}=i( Q_- \hee{+} + Q_+ \hee{-})~,\nn\\ 
\dq{3}=i( Q_- \hcc{+} - Q_+ \hcc{-}) ~.
\end{eqnarray}
The decomposition of the $\nZZ$ invariants $W,B$ is
\begin{equation}
\label{eq::basis}
F^i = \left.\left( \begin{array}{c}
W \\
B \\
\bar{W} \\
\bar{B} \\
\tilde{W} \\
\tilde{B} \\
\bar{\tilde{W}}\\
\bar{\tilde{B}}
\end{array}
\right) \right|= \frac{1}{2}\left( \begin{array}{rrrrrrrr}
-i&-i& 1& 1& 0& 1& 0& i\\
 i&-i&-1& 1& 1& 0& i& 0\\
 i& i& 1& 1& 0& 1& 0&-i\\
-i& i&-1& 1& 1& 0&-i& 0\\
-i&-i&-1& 1&-1& 0& 0&-i\\
 i&-i& 1& 1& 0&-1&-i& 0\\
 i& i&-1& 1&-1& 0& 0& i\\
-i& i& 1& 1& 0&-1& i& 0
\end{array} \right) \left( \begin{array}{c}
iD_+ \hee{-}\\
iD_- \hee{+}\\
iD_+ \hcc{-}\\
iD_- \hcc{+}\\
\dq{1}\\
\dq{2}\\
\dq{3}\\
f
\end{array}
\right)~~.
\end{equation}

\subsubsection{Matter couplings in $\nll$ superspace}
We start from the gauged $\nZZ$ Lagrange density:
\begin{equation}
\label{eq::chiral_gauged_action} 
K_\phi \left( \phi + \bar{\phi} + 
V^\phi,\chi + \bar{\chi}+ V^\chi, i(\phi - \bar{\phi} + 
\chi - \bar{\chi}) + V^\prime \right)~.
\end{equation}
We reduce to $\nll$ superfields, which in the Wess-Zumino
gauge 
\beq
V^\phi|=0~~,~~~V^\chi|=0~~,~~~V'|=0~~,
\eeq{largewz}
are simply
\begin{eqnarray}
\phi|\aleq\phi\nn~,\\
\chi|\aleq\chi\nn~,\\
(Q_+ \hph)| \aleq +iD_+ \hph - (\hee++i\hcc+) - A_+\nn~,\\
(Q_+ \hch)| \aleq +iD_+ \hch -  (\hee++i\hcc+) + A_+ \nn~,\\
(Q_- \hph)| \aleq +iD_- \hph - (\hee-+i\hcc-) - A_- \nn~,\\
(Q_- \hch)| \aleq -iD_- \hch +(\hee--i\hcc-)  - A_- ~.
\end{eqnarray}
It is useful to introduce the notation
\begin{eqnarray}
\hPh{i} \aleq (\hph,\hphb,\hch,\hchb)
\end{eqnarray}
and the covariant derivatives
\begin{equation}
\hcd{\pm} \hPh{i} = D_\pm \hPh{i} + A_\pm k^i~.
\end{equation}
This gives
\begin{equation}
Q_\pm \hPh{i} = \jj{\pm}{i}{j} \hcd{\pm} \hPh{j} + \hee{\pm} \jj{\mp}{i}{j} k^j + \hcc{\pm} \Pi^i{}_j k^j 
\end{equation}
and
\begin{eqnarray}
2 Q_+ Q_- \hPh{i} =\!\! \!\!\!\!\! && D_+ (\Pi^i{}_j\hcd{-}\hPh{j} -\hee{-}k^i 
- 2 \hcc{-}\jj{-}{i}{j}k^j )\nn\\ 
 && - D_- (\Pi^i{}_j\hcd{+}\hPh{j} -\hee{+}k^i - 
2 \hcc{+}\jj{+}{i}{j}k^j ) + 2 \tilde L^i{}_\alpha \dq{\alpha}
\end{eqnarray}
where $\alpha=1,2,3$ and
\begin{equation}
\tilde L = -\frac{i}{2}\left( \begin{array}{rrr}
2&0&i\\
2&0&-i\\
0&2&i\\
0&2&-i\\
\end{array}
\right)
\end{equation}
The $\nll$ superspace Lagrange density is (after integrating by parts and using the
isometry)
\ber
\mathcal{L}=K_{ij}\!\left[\begin{array}{l}
- \ft{1}{2}\left( \hcd{+}\hPh{i} \left( \Pi^j{}_l \hcd{-}\hPh{l} - 2 \hcc{-}\jj{-}{j}{l}k^l \right)
+ \left(\Pi^i{}_k \hcd{+}\hPh{k} - 2 \hcc{+} \jj{+}{i}{k}k^k \right) \hcd{-}\hPh{j} \right) 
\\[1mm]
+ \left( \redQp{i}{k} \right)\!\left( \redQm{j}{l} \right) \end{array}\!\!\right]&&\nn\\
 + \,\,K_i \tilde L^i{}_\alpha \dq{\alpha}~.\qquad&&
\eer{eq:matterac}
The large vector multiplet has the gauge-invariant spinors $\Xi_\pm^A$; it is useful
to isolate their contribution to expose the underlying $\nll$ 
gauged nonlinear $\sigma$-model.
We define the matrices:
\begin{eqnarray}
E_{kl} \aleq \ft{1}{2} K_{ij} \left( 2\jj{+}{i}{k}\jj{-}{j}{l} - \Pi^i{}_k \delta^j{}_l - \Pi^j{}_l \delta^i{}_k  \right) \\[2mm]
E_{Al} \aleq \left( \begin{array}{c}
 K_{ij} \jj{-}{i}{k} k^k \jj{-}{j}{l} \\[1mm]
 K_{ij} \left( \jj{+}{i}{k} k^k \delta^j{}_l + \Pi^i{}_k k^k \jj{-}{j}{l} \right)
\end{array}
\right) \\[2mm]
E_{kA} \aleq \Big( K_{ij} \jj{+}{i}{k}\jj{+}{j}{l}k^l ~,~ 
K_{ij}\! \left( \jj{-}{j}{l}k^l \delta^i{}_k + \jj{+}{i}{k} \Pi^j{}_l k^l \right) \Big) \\[2mm]
E_{AB} \aleq \left( \begin{array}{cc}
K_{ij} \jj{-}{i}{k} k^k \jj{+}{j}{l} k^l & K_{ij} \Pi^i{}_k k^k \jj{+}{j}{l} k^l \\[1mm]
K_{ij} \jj{-}{i}{k} k^k \Pi^j{}_l k^l & K_{ij} \Pi^i{}_k k^k \Pi^j{}_l k^l
\end{array} \right)
\end{eqnarray}
We find
\begin{eqnarray}
\mathcal{L}= \left( \Xi_+^A + \hcd{+}\hPh{i} E_{iC} E^{CA} \right) 
E_{AB} \left( \Xi_-^B + E^{BD} E_{Dj} \hcd{-}\hPh{j}  \right) \nn\\[1mm]
+ \hcd{+}\hPh{i} \left( E_{ij} - E_{iA} E^{AB} E_{Bj}\right) \hcd{-} 
\hPh{j} + K_i \tilde L^i{}_\alpha \dq{\alpha}
\end{eqnarray}
with $E^{AB}$ the inverse of $E_{AB}$.
\subsubsection{The vector multiplet action}
A general $\nZZ$ action for the large multiplet can be written as
\begin{equation}
S_a = \int d^2 \xi D_+ D_- Q_+ Q_- \left( F^i F^j g_{ij} + \bbG{+}^A \bbG{-}^B m_{AB} \right)~,
\end{equation}
where the ranges for indices are $i,j = 1, \cdots, 8~;~AB = 1,2$, and the spinor invariants were arranged into tuples
\begin{equation}
\bbG{\pm}^A = (\bbG{\pm},\bbGB{\pm})~.
\end{equation}
Other terms of the type $(\bbD{\pm} ,\bbDB{\pm} )(\bbG{\pm},\bbGB{\pm})$  could be integrated by parts to give the $W$ and $B$ invariants. One could also add superpotential terms.

This action can be reduced to $\nll$ using the block-(twisted)chirality of $F$ and the semichirality of $\mathbb{G}$. In general, one finds terms with higher derivatives; it does not
seem possible to find a sensible kinetic action, but we leave a complete analysis for future work.

\subsubsection{Linear terms}
As discussed above for the semichiral vector multiplet, linear couplings of unconstrained 
Lagrange multiplier fields multiplying the field-strengths are needed to discuss T-duality.
In $\nZZ$ superspace, we constrain the field-strengths $\bbG{\pm}$ to vanish with unconstrained
complex spinor Lagrange multiplier superfields $\Psi_\mp$:
\beq
\mathcal{L}_{linear}= i\left(\Psi_+\bbG-+\Psi_-\bbG+
+\bar\Psi_+\bbGB-+\bar\Psi_-\bbGB-\right)~.
\eeq{largevlin}
When we integrate by parts and define semichiral Lagrange multpliers 
$\bbX{L,R}=-i\bbDB\pm\Psi_\mp$, we find
\begin{equation}
\mathcal{L}_{linear}= \bbX{L} V + \bbXB{L} \bar{V} + 
\bbX{R} \tilde{V}+ \bbXB{R} \bar{\tilde{V}}~.
\end{equation}
Reducing to $\nll$ supperspace, and defining $\nll$-components for the Lagrange multipliers
as in (\ref{semixdef},\ref{semipsidef}) we find
\begin{eqnarray}
\mathcal{L}_{linear} =\!\!\!\!\!\!\!&&
\psi_- \left( i \hee{+} - \hcc{+} \right) +\ft12 X_L 
\left( (\dq{2}+\dq{1}) +i(f+\dq{3}) \right) \nn \\
&&+\,\, \bar{\psi}_- \left( - i\hee{+}-\hcc{+}  \right) 
+ \ft12\bar{X}_L \left( -  (\dq{2}+\dq{1}) +i (f+\dq{3}) \right) \nn \\
&&+ \,\,\psi_+ \left( - i \hee{-} + \hcc{-} \right) 
+\ft12 X_R \left( -(\dq{2}-\dq{1}) -i (f-\dq{1})\right) \nn \\
&&+ \,\,\bar{\psi}_+ \left( i \hee{+} -i \hcc{+}  \right) 
+ \ft12\bar{X}_R \left(  (\dq{2}-\dq{1}) -i (f-\dq{1}) \right)~.
\end{eqnarray}
We can easily integrate out $\psi_\pm$ and their complex 
conjugates; this  $\Xi_\pm^A$ from the action.
We are then left with the usual T-duality transformation 
as we shall discuss in \cite{LRRvUZ}. 

\bigskip
\noindent{\bf\large Note}:

\noindent 
As we were completing our work, we became aware of
related work by S.J.~Gates and W.~Merrell; we thank them for agreeing
to delay their work and post simultaneously.
\bigskip\bigskip

\noindent{\bf\Large Acknowledgement}:
\bigskip

\noindent  
UL supported by EU grant (Superstring theory)
MRTN-2004-512194 and VR grant 621-2006-3365.
The work of MR and IR was supported in part by NSF grant no.~PHY-0354776. 
The research of R.v.U. was supported by 
Czech ministry of education contract No.~MSM0021622409. 
The research of M.Z. was
supported by VR-grant 621-2004-3177.

\end{document}